\documentstyle[11pt]{article}
\begin{document}

\renewcommand{\thefootnote}{*}

\baselineskip=16.5pt plus 0.2pt minus 0.2pt
\vsize=22cm

\pagestyle{myheadings}
\markboth{Draft }{\today $\;\;\;$ { \sf }}

\title{\vspace*{-2.5cm}\hspace*{\fill}{\normalsize LA-UR-97-4374} \\[1.5ex]
\vspace*{0.5cm}
Relativistic flows on a spacetime lattice}
\vspace{1cm}
\author{N.L. Balazs$^1$\thanks{E. Mail: balazs@sbhep.physics.sunysb.edu}{\ },
B.R. Schlei$^{2,3}$\thanks{E. Mail: schlei@LANL.gov}{\ }, and
D. Strottman$^2$\thanks{E. Mail: dds@LANL.gov}{\ }{\ }
\\[1.5ex]
\hspace*{-1.1cm}{\it $^1$ Department of Physics and Astronomy, 
SUNY at Stony Brook, NY 11794}\\ 
\hspace*{-1.1cm}{\it $^2$ Theoretical Division, DDT-DO, Los Alamos National 
Laboratory, NM 87545}\\ 
\hspace*{-1.1cm}{\it $^3$ Physics Division, P-25, Los Alamos National 
Laboratory, NM 87545}
}


\date{\today}\maketitle

\begin{abstract}
The relativistic extension of non-relativistic hydrodynamics suffers from
notorious difficulties.  In non-relativistic hydrodynamics where
difficulties also abound, it has proved a useful supplement to study
lattice models which can imitate viscous fluid flow.  In this paper we
construct a relativistic spacetime lattice and construct a dynamics of
points, thus a relativistic cellular automaton over it, to model
relativistic fluid flow.  A simple example is also explicitly studied,
and some numerical results with figures are shown in the last section.
\vspace{.3cm}
\end{abstract}


\section{INTRODUCTION}

A causal description of matter under extreme conditions is a very
difficult task.  A relativistic description of heavy ion physics suffers
strongly from this predicament.  A {\sl microscopic} causal description
is not available since we do not know sufficiently well the microscopic
interactions among the constituents under these conditions.  A proper
relativistic interaction must be local and handled through fields.
Usually, however, one attempts to formulate a theory entirely in terms of
the matter degrees of freedom.

At the same time it is not ensured that a more {\sl macroscopic, causal
and local description of nuclear matter alone,} and in which the field
degrees of freedom responsible for the interaction are entirely removed,
exists.  According to Hilbert\cite{Hilb}, Bogoliubov\cite{Bog} and
others, such a separation and elimination of the degrees of freedom would
require the coexistence of widely disparate time scales in the evolution
of the system; one associated with the field degrees of freedom and the
other with the matter degrees of freedom.  The existence of different time
scales in the field degrees of freedom and the matter degrees of freedom
is not at all assured in the highly relativistic region! (Such a
separation of the fields from the matter was given by van Kampen \cite
{Kamp7} for a highly simplified, non-relativistic model and resulted in
integral equations.  For matter interacting through Boson fields, see de
Groot {\it et al}.\cite {Groot1980}, p.79.)

If, however, we disregard these worries, we may attempt to use the
simplest macroscopic causal description of the matter (with the exclusion
of the fields), namely, hydrodynamics, supported with local
thermodynamical relations, both adapted for extreme conditions, {\sl
i.e., for high velocities and large internal energies}. Such forms of
relativistic hydrodynamics have been known for a very long time
\cite{Eck,Land}.

The tool of relativistic hydrodynamics has, alas, its own difficulties.
In perfect fluid dynamics the mean free path in a fluid is zero.  The
introduction of dissipation (viscosity and heat conduction) provides for
a non-zero mean free path.  At relativistic energies where transparency
effects become important, the introduction of such effects would appear
essential.  However, it is unclear how one can separate the dissipative
effects from the others at relativistic energies.  This separability also
hinges on the existence of different time scales in the motion of
relativistic matter alone.  To uncover such a possibility one should
translate Bogoliubov's original ideas in an invariant manner to a
relativistic description.  This has not yet been done.

With dissipation, the Euler equations generalize, albeit with ambiguities,
to something much more complicated.  (A review of the basic problems with
detailed literature references can be found in ref. \cite{Balr}.) Two of
the ambiguities are:

1) A result of introducing dissipation is that the energy flow will no
longer be in the same direction as the matter flow or the entropy flow.
One has then a dilemma in choosing a rest frame: a frame with zero energy
flow may have a nonzero matter flow, and nonzero entropy flow.  This
ambiguity has led to different choices: Eckart chose a frame in which the
matter flux was zero, Landau one in which the energy flux vanishes.  The
third natural choice, the vanishing of the entropy flux, could be
exploited with equal justification.  Linear combinations of these fluxes
could also be used.  The resulting theories are not equivalent except in
the limit of zero dissipation.  This makes one wonder why one should
introduce a rest frame at all! However, near equilibrium the
instantaneous local rest frame is needed to specify the thermodynamical
quantities which appear in the fluid equations.

2) Instabilities and acausal behavior may arise as demonstrated by
Hiscock and Lindblom\cite{Hiscock1985} and by Geroch and
Lindblom\cite{Geroch1990,Geroch1991}.  In the Eckart frame all solutions
have instabilities and are acausal.  The instabilities grow exponentially
with a time scale on the order of 10$^{-25}$ seconds. In the Landau frame
almost all solutions are unstable.  A recent theory espoused by
Carter\cite{Carter1989} has been found to have similar
problems\cite{Olson}.

An alternative approach that attempts to include dissipation has been
advanced by Israel and Stewart\cite{Is1979,Is1988,St71,St77} and is a
generalization of a nonrelativistic theory by M\"uller\cite{MU67}.
However, the five Euler equations are replaced in the Israel and Stewart
model by fourteen equations and the three coefficients of viscosity and
the coefficient of conduction in the Navier-Stokes equation is replaced by
six new coefficients.  Hiscock and Lindblom\cite{Hiscock1983,Hiscock1988}
have also analyzed Israel's theory. In the Eckart frame instabilities
still persist, but in the Landau frame, solutions are stable and causal
for modest deviations from equilibrium; for large deviations from
equilibrium and for large values of internal energy, problems persist.  A
relativistic kinetic theory for dilute gases has also been developed,
starting with the work of Synge \cite{Synge} and developed greatly by
Israel and others.  (See the detailed discussion in the book by de Groot
et al.\cite{Groot1980} and in the articles by
Israel\cite{Israel1972,Israel1989}). However, not all the difficulties and
ambiguities have been successfully resolved.

A broad class of relativistic dissipative theories have been reviewed by
Geroch
\cite{Geroch} who concludes that there are a vast number of competing
theories whose physical usefulness is questionable in the regimes where
they agree with each other, while their differences are important only in
those regimes in which they already break down.

Past experience on non-relativistic hydrodynamics
\cite{Doolen1990,Doolen1991} suggests that a clearer understanding of
many features of the fluid flow may be gleaned through the use of
discrete models, lattice automata.  In this note we describe how such
lattice automata can be constructed in the relativistic domain.  Since in
these lattice automata the interactions are not handled through fields,
but through contact interactions, the separation of the matter degrees of
freedom from those of the field is not needed.

\section{SPACETIME LATTICES}

Non-relativistic lattice automata make use of the {\it Galilean} group.
The Galilean group is the product of the Euclidean group (rotations and
translations in space) and the (one dimensional) translation group along
the time axis.  Since the translations along time are independent of the
spatial transformations, the structure of the lattice is entirely
determined by {\sl the discrete subgroups of the Euclidean group}, the
standard crystallographic groups of the required spatial dimensions.  The
incorporation of the time translations are unnecessary.  A space-like
vector connects the lattice sites in Euclidean space, and its components
transform corresponding to contravariant vectors under the action of the
discrete rotations.  Because the hopping speed from lattice point to
lattice point is not restricted by any invariance requirement there will
always be a Galilean transformation which can reduce {\sl any} hopping
velocity to rest.

The incorporation of the conservation laws in a non-relativistic lattice
model simply means that the hopping rules must be such that a) if a point
hops without encountering another one its velocity is unchanged; b) if
two or more points meet, the rule determining the next step must satisfy
the conservation laws.  As the time translations play no role, the
updating rules of a {\sl Newtonian automaton need to be concerned only
with the crystallographic lattice}.

The situation changes if the automaton is relativistic; now a full
spacetime lattice must be used.  As before, invariance requirements
enter in the same two places, but with different consequences:

a) the spacetime lattice must now be invariant on {\sl a discrete subgroup
of the Poincar\'{e} group}, the invariance group of relativistic dynamics;

b) the hopping rules must incorporate the {\sl relativistic}
energy-momentum conservation laws.  \\

The {\it Poincar\'{e}} group is formed by Lorentz transformations and
spacetime translations.  It has an infinite number of discrete subgroups,
producing an infinite number of different spacetime lattices, in
contradistinction to the finite number of crystallographic lattices.  The
microscopic dynamics will consist of hopping on such a spacetime lattice
according to certain rules.  The discrete Lorentz transformations
used in constructing such a lattice will link the space and time
variables. As a consequence, not {\sl all} frame velocities are permitted
but only those which can be reached by the class of Lorentz
transformations present in the discrete subgroup envisaged.  {\sl
Accordingly, one can preserve only as much of full Lorentz invariance} (
generated by the continuous group) {\sl as allowed by the discretization.} If
we also desire - as we shall do - that any {\sl microscopic} one-step
spacetime hop velocity could be transformed away, then only those hop
velocities are permitted which are generated through boosts produced by
the allowed class of discrete Lorentz transformations. (This condition
could conceivably be relaxed, arguing that a local ``rest'' frame means
the vanishing of some {\sl average} flux, and not a microscopic velocity.  For
example, the instantaneous speed of a Dirac electron is always the speed
of light, due to its Zitterbewegung, while its average velocity can be
anything less than c.)

The actual construction of the hopping rules require the construction
of the spacetime lattice in order to determine the allowed Lorentz
transformations and thereby the allowed world momenta.  To do this we
will study a specific example in greater detail.  This simple example is
too primitive to be physically realistic, but it will exhibit explicitly
the methods embodied.

 An interesting previous attempt has been made without the use of the above
requirements by Hersbach\cite{hersbach}. There a one dimensional model was
obtained in which the
particles move right or left with the velocity of light.  This has serious
consequences: a) there
is no rest frame for the microscopic particle motion, b) in the collisions
detailed balancing and
time reversal invariance are violated.

\section{THE CONSTRUCTION OF SPACETIME LATTICES }

We have previously specifed \cite{Balazs} that the construction advocated
here fails in a (1+1)
dimensional spacetime, but not in higher dimensions.  Therefore we
will describe the general approach in (3+1)  and then in
 (2+1) spacetime dimensions. The actual transport model will use the latter.

 An inertial frame specifies a
coordinate system with straight axes along Minkowski (or M) orthogonal
frame vectors
\(\mbox{\bf e}(a), a = 0,1, 2, 3\), satisfying the conditions
\begin{equation}
\mbox{\bf e}(a)\cdot\mbox{\bf e}(b) = \eta (a,b),
\end{equation} where $\eta (a,b)$ is given by
\begin{equation}
 \eta (a,b) = \left(\matrix{
-1 & 0 & 0 & 0 \cr
0 & 1 & 0 & 0 \cr
0 & 0 & 1 & 0 \cr
0 & 0 & 0 & 1 \cr
}\right) . \end{equation}

Since this spacetime is flat we can introduce a finite size radius vector
\(\mbox{\bf r}\) connecting a point in spacetime with the origin.  This
point has the coordinates $x^a$ where
\begin{equation}
 \mbox{\bf r} = \sum x^a \mbox{\bf e}(a).  \end{equation}
One can similarly introduce an arbitrary world vector \(\mbox{\bf u}\),
expressed as a linear combination of frame vectors, the coefficients
being the contravariant components $u^a$,
\begin{equation}
 \mbox{\bf u} = \sum u^a \mbox{\bf e}(a).  \end{equation} (It is useful to
stress that changing frames means two things: changing the {\it frame
vectors }\(\mbox{\bf e}(a), a = 0,1, 2, 3\), and the associated {\it
components} of the particular vectors. The components transform as
contravariant components, while the frame {\it vectors} change as if
their names,{\it the labels $(a)$} were covariant indices.)

The scalar product of two world vectors \(\mbox{\bf u}\) and \(\mbox{\bf
v}\) is immediately given through their contravariant components as
\begin{equation}
\mbox{\bf u}\cdot\mbox{\bf v} = - u^0 v^0 + u^1 v^1 + u^2 v^2 + u^3 v^3.
\end{equation} Our notation will be slightly inconsistent, denoting
sometimes a world vector in bold face as \(\mbox{\bf u}\), or by its
components in a frame as $u^{i}$, or just by $u$. Where this would give
rise to any ambiguity, we shall be precise.

The simplest lattices are cubic.  Such a lattice can be constructed by
selecting {\sl lattice points} obtained through discrete translations of
unit step-length along the coordinate axes \cite{Balazs}.   Then a general
lattice point, indexed by
the integers
$(k,l,m,n)$, will have a radius vector {\bf r}
\begin {equation}
\mbox{\bf r} = k\mbox{\bf e}(0) + l\mbox{\bf e}(1) + m\mbox{\bf e}(2)
+n\mbox{\bf e}(3).
\end{equation} Letting $(k,l,m,n)$ run through all positive and negative
integers, including zero, a spacetime lattice is generated.

When will the lattice sites be unchanged under Lorentz transformations?
Consider a Lorentz transformation matrix with integer coefficients.  The
original quadruplet of integers $(k,l,m,n,)$ transforms into another
quadruplet of integers
$(k^{'},l^{'},m^{'},n^{'})$.  Thus, from the active point of view the
transformation shifts a lattice point into another lattice point;
from the passive point of view it relabels a fixed lattice point.  Thus
the existence of an invariant spacetime lattice depends on the existence
of Lorentz transformations specified by matrices with {\sl integer}
elements with respect to a particular set of basis vectors.

Let
\begin{equation}
 L = \left(\matrix{
a^0 & b^0 & c^0 & d^0 \cr
a^1 & b^1 & c^1 & d^1 \cr
a^2 & b^2 & c^2 & d^2 \cr
a^3 & b^3 & c^3 & d^3 \cr
}\right) \end{equation}
be a Lorentz matrix specified in an inertial frame with frame vectors

$\begin{array}{cccc}
e(0)=\left(\begin{array}{c}1\\0\\0\\0\end{array}\right),&
e(1) =\left(\begin{array}{c}0\\1\\0\\0\end{array}\right),&
e(2) = \left(\begin{array}{c}0\\0\\1\\0\end{array}\right),&
e(3) = \left(\begin{array}{c}0\\0\\0\\1\end{array}\right). \end{array}$\\
Applying $L$ to these frame vectors we generate a new frame with the new frame
vectors

$\begin{array}{cccc}
e^{'}(0) =
\left(\begin{array}{c}a^{0}\\a^{1}\\a^{2}\\a^{3}\end{array}\right),&
e^{'}(1) = \left(\begin{array}{c}b^{0}\\b^{1}\\b^{2}\\b^{3}\end{array}\right),&
e^{'}(2) = \left(\begin{array}{c}c^{0}\\c^{1}\\c^{2}\\c^{3}\end{array}\right),&
e^{'}(3) =
\left(\begin{array}{c}d^{0}\\d^{1}\\d^{2}\\d^{3}\end{array}\right).
\end{array}
$

Hence the new frame vectors are the vectors \(\mbox{\bf a, b, c, d}\) with
components composed from the columns of the matrix.  Since a Lorentz
transformation must preserve M orthonormality, the following conditions
are imposed on the matrix elements, expressed as conditions on the frame
vectors:
\begin{eqnarray}
 - \mbox{\bf a}\cdot\mbox{\bf a} & = & \mbox{\bf b}\cdot\mbox{\bf b} =
\mbox{\bf c}\cdot\mbox{\bf c} = \mbox{\bf d}\cdot\mbox{\bf d} = 1\\
 \mbox{\bf a}\cdot\mbox{\bf b}& = &\mbox{\bf a}\cdot\mbox{\bf c} =
\mbox{\bf a}\cdot\mbox{\bf d} = \nonumber\\
\mbox{\bf b}\cdot\mbox{\bf c}& = & \mbox{\bf b}\cdot\mbox{\bf d} =
\nonumber\\
\mbox{\bf c}\cdot\mbox{\bf d}& = & 0. \end{eqnarray}

These conditions lead to Diophantine equations.

In two dimensions there are no solutions unless we use null coordinates;
in three dimensions -- to which we now restrict the discussion -- there
are an infinity of solutions as indeed there are also in four
dimensions.  The one containing the smallest positive integers is given
by the matrix
\begin{equation}
 L(1) = \left(\matrix{
3 & 2 & 2 & \cr
2 & 1 & 2 & \cr
2 & 2 & 1 & \cr
}\right) ,
\end{equation}
where the columns specify the frame vector components of the moving frame.
(Other solutions can be found in \cite {Balazs,new}).  The first column
\nobreak corresponds to a world velocity with components
$(1/\sqrt{1 - v^2}, v_1/\sqrt{1 - v^2}, v_2/\sqrt{1 - v^2})$, hence, in
the example above, to a velocity $\sqrt{8}/3$ in units of the velocity of
light.

How are the time and length units to be obtained in the moving frame? In
the conventional case the marks denoting the units on the different
coordinate axes are specified by the intersection of the coordinate axes
with the unit calibration - hyperboloids.  The time axis intercepts the
hyperboloid of two sheets, while the two space-like axes intercept the
hyperboloid of one sheet.  In the present case there is a lattice point
at each of these intersections, and these lattice points are the nearest
to the origin located on the coordinate axes.  Thus the conventional
specification of space and time units is the same as taking the {\sl
lattice point separation }along each axis as the unit.  (One may view the
construction of the spacetime lattice as seeking all those points where
the two length specifications coincide, the one through the calibration
hyperboloids, and the other through spacing equally the lattice points on
a coordinate axis, sliding them like beads.)

From $L(1)$ we can generate three other Lorentz transformations satisfying
all conditions, Eqs. (8) and (9), by rotating the original spatial axes.
This results in four $L$ transformations altogether; these are \vspace
{0.1in}\\
\vspace {0.1in}
$ L(1) = \left(\begin{array}{rrr}\;\;\;3 & \;\;\;2 & \;\;\;2 \\
 2 &  1 &  2 \\
 2 & 2 & 1 \end{array}\right) ,\\ \vspace {0.1in}
L(2) = \left(\begin{array}{rrr} 3 & -2 & 2 \\
   -2 & 1 & -2 \\
   2 & -2 & 1 \end{array}\right),\\ \vspace {0.1in}
L(3)\ = \left(\begin{array}{rrr} 3 & -2 & -2 \\
   -2 & 1 & 2 \\
   -2 & 2 & 1 \end{array}\right),\\ \vspace {0.1in}
L(4)\ = \left(\begin{array}{rrr} 3 & 2 & -2 \\
   2 & 1 & -2 \\
   -2 & -2 & 1 \end{array}\right).  \\\vspace {0.1in}$
Our model will be built on these four transformations.

\section{DYNAMICAL QUANTITIES}

 1) Invariance in this lattice is a {\sl reduced} relativistic invariance.
It shall mean invariance under transformations generated by the four
discrete Lorentz transformations constructed above, their powers and their
products.

2) In this frame the components of the four possible hopping world
velocities, $u_A,\; ( A
= 1,2,3,4;$ modulo $4$) are given by the first column of each of these
matrices. (Observe that capital letters refer to the name of the
particular world velocity, and not to components.)

$\begin{array}{cccc} u_1
=\left(\begin{array}{r}3\\2\\2\end{array}\right),& u_2
=\left(\begin{array}{r}3\\-2\\2\end{array}\right),& u_3 =
\left(\begin{array}{r}3\\-2\\-2\end{array}\right),& u_4 =
\left(\begin{array}{r}3\\2\\-2\end{array}\right). \end{array}
$\\
Thus in this frame we find the following values for the dynamical
quantities associated with a particle: energy = $3mc^{2}$, speed =
$(\sqrt{8}/3) c$, magnitude of momentum =
$\sqrt{8} mc$. This is a very relativistic speed indeed; all other
solutions have even higher speeds. (We have chosen these world velocities
as fundamental, since
any one of them can be reduced to rest  by using one of the basic Lorentz
transformations.)

For the present case a sort of one particle (or reduced) phase space
lattice can be constructed.  (It is not a {\it bona fide} phase space
since there is no symplectic form associated with this discrete
manifold.) To each spacetime lattice point we attach as arrows to the
vertices, $u_{A}, A = 1,2,3,4$.  Their tips will lie on the upper sheet
of unit hyperboloids hovering over each lattice point.  A microscopic
dynamical state of a particle is specified by giving the location of the
lattice point (to fix its spacetime position), and the tip of one of the
attached $u_{A}\:'s$.  The evolution is now conceived by hopping from tip
to tip.

Three comments are in order.  First, the previous considerations can be
extended to a four dimensional spacetime.  An infinity of solutions exist
and are reported elsewhere \cite{new}.  Second, more complicated spacetime
lattices can be generated that will increase isotropization in spacetime,
and hence, also in the lattice points on the space-like cuts.  Third, we
can increase the isotropization in the world velocities (or world
momenta) by applying one or more allowed Lorentz transformations to any
member of our original $u_A$ set and interpreting the generated time-like
unit vectors as additional world velocities in the original frame
(instead as the original vector in a new frame). (These new world
velocities still can be reduced
to rest applying products of our basic Lorentz transformations.)  In the
phase space picture this
would produce a larger set of arrows attached to each lattice point with
all tips touching each unit
hyperboloid.

For these reasons the following Sections will be phrased in more general
terms and not restricted to the special case described above.  However,
where appropriate we shall point out if the present model exhibits
unexpected, special or odd features.

\section{MICRODYNAMICS}

Lattice points can be populated by mass points having one of the four
world velocities; these mass points must jump to another lattice point at
each iteration. The units used here make the mass and the speed of light
unity; thus, the world velocity is numerically equal to the world
momentum.  (Notice that here the variable labeling the evolution is the
invariant iteration number $q$, and {\sl not} the time, which is one of
the frame-dependent labels of $r$.  In the future we shall specify the
location and iteration number together, as ${r,q}$.  The present simple
model has the special feature that in this particular frame, each
iteration advances the time by three units since the time component of
each
$u_A$ is 3.)

Let the scalar quantity $n_{A}(r,q)$ be unity if the lattice point $r$ is
occupied at the q'th iteration by a point with world velocity $ u_{A}$,
and zero otherwise.  The evolution equation of the four discrete
functions is given by
\begin{equation}
 n_{A}(r + u_{A},q + 1) = n_{A}(r,q) + c_{A}(r,q).  \end{equation}
The
first term on the right describes pure streaming.  The collision term,
$c_{A}(r,q)$, determines the state of affairs at $r + u_{A},q+1$ arising
from the collision at $r,q$.  The collision must satisfy the conservation
of world momenta. In the present model this is enforced as follows.  (For
the non-relativistic situation see ref. \cite{Frisch}.) Only two points
can collide, and only if the collision is head on, and only if it results
in a rebound with the spatial components of the incoming world velocities
turned ninety degrees (in this special frame!).  Thus,
$c_{A}(r,q)$ is equal to the configurations producing a $Gain$ minus the
configurations producing a $Loss$ at $r + u_{A},q+1$.  $Loss$ occurs when
at $r$ the configurations with $u_{A},u_{A+2}$ are occupied to produce a
head-on collision that will remove the point with $u_{A}$.  This can only
happen if the configurations with $u_{A+1},u_{A+3}$ are both unoccupied,
to receive the points with the turned incoming world velocities.  $Gain$
occurs when at $r,q$ the configurations with
$u_{A+1},u_{A+3}$ are occupied, and $u_{A},u_{A+2}$ are unoccupied.  This
gives
\begin{eqnarray}
 Loss &=& n_{A}n_{A+2}(1 - n_{A+1})(1 - n_{A+3}),\nopagebreak\\
 Gain &=& n_{A+1}n_{A+3}(1 - n_{A})(1 - n_{A+2}), \\ c_{A}(n)&=& Gain -
Loss \nonumber\\
 &=& n_{A+1}n_{A+3}(1 - n_{A})(1 - n_{A+2}) - \nonumber\\
& & n_{A}n_{A+2}(1 - n_{A+1})(1 - n_{A+3}). \end{eqnarray}
(All quantities are evaluated at $r,q$.)

It is an easy exercise to show that the two conservation laws of particle
number and world momentum
\begin{eqnarray}
\sum_{A} n_{A} c_{A}(n) &= &0,\\
\sum_{A} u^{i}_{A}n_{A}c_{A}(n)&=&0,
\end{eqnarray} are satisfied.  The solution, $n_{A}(r,q)$, gives the most
detailed microscopic description of the system at any iteration.  Having
$n_{A}(r,q)$, all relevant quantities at $r,q$ can be
directly evaluated.  These are the

\begin{eqnarray}
\rm {number\: of\: particles } &=& \sum_{A} n_{A},\\
\rm {total\: world \:momentum } &=&\sum_{A} u^{i}_{A}n_{A},\\
\rm {microscopic\: energy\: momentum\: tensor} &=& \sum_{A}
u^{i}_{A}u^{k}_{A}n_{A} . \end{eqnarray}

\section {KINETIC THEORY}

A different description is provided by the precepts of kinetic theory
using a probabilistic approach.  We can average $n_{A}$ in many different
ways, for example, averaging over different initial data, or averaging
over spacetime regions of various sorts, or assigning {\it a priori}
probabilities to the occurrence of the initial
$n_{A}$'s, etc.  One arrives at four distribution functions
$f_{A}(r,q)$, one for each of the four world velocities, that specify the
probability of finding a point with world velocity $u_{A}$ at the lattice
point $r$ at a given iteration number $q$.  Thus, the distribution
functions are attached to {\sl lattice points } in the discrete phase
space, and consequently they are {\sl not densities} in a phase space as
in the usual continuum theories.

What are the evolution equations of these functions? If (following
Boltzmann) we neglect correlations between the different $n_{A}$'s, and
average the evolution equations of the $n_{A}$'s, these equations become
identical in form with the evolution equations for the
$n_{A}$'s, simply replacing the latter by the corresponding $f_{A}$'s,
{\sl i.e.,}
\begin{equation}
 f_{A}(r + u_{A},q + 1) = f_{A}(r,q) + c_{A}(r,q).  \end{equation} where
\begin{eqnarray}
 c_{A} &= f_{A+1}f_{A+3}(1 - f_{A})(1 - f_{A+2})\\
 & - f_{A}f_{A+2}(1-f_{A+1})(1 - f_{A+3}). \end{eqnarray} Just as for
the $n_{A}$, here also the two conservation laws
\begin{eqnarray}
\sum_{A} f_{A}c_{A}(f)&=& 0,\\
\sum_{A} u^{i}_{A}f_{A}c_{A}(f)&=& 0,
\end{eqnarray}
are satisfied.  The four coupled equations in Eq. $21$ correspond to the
relativistic Boltzmann equation.  (There are four equations only, since
in our model there are only four world momenta, $A = 1,2,3,4$ present
instead of a continuous range.  In general there will be as many
equations as there are allowed world velocities.)

Given the starting data $f_{A}(r,0)$ at the zeroth iteration, these
equations give us
$f_{A}(r,q)$, the probability finding a particle with world velocity
$u_{A}$ at $r,q$. Observe that the starting data refer to the zeroth
iteration and not to time equal zero. The latter is a frame-dependent
notion.  It may happen, however, that the averaging process itself
tacitly introduces an initial frame dependence; for example, averaging
over initial data on a space-like, $t = 0$, plane.  This often corresponds
to the actual physical situation where experiments are performed with the
same setup at different times in the laboratory frame, and assuming that
the setup is the same at each starting time.

The relevant macroscopic, or mean quantities at $r,q$ are as follows.
\begin{eqnarray}
\rm{density:}\: \rho &=& \sum_{A} f_{A}\\
\rm{world\: momentum \:flow,\: or\: particle \:flow:}\: {\it N^{i}} &=&
\sum_A u^{i}_{A}f_{A}\\
\rm{ mean\: entropy\: flow:\:} {\it S^{i}} &=& \sum_A u^{i}_{A}
[f_{A}logf_{A} + \nonumber\\&      &(1 - f_{A})log (1 - f_{A})]\;\;\\
\rm{ energy\: momentum\: tensor:\:} {\it T^{ik}} &=& \sum_{A} u^{i}_{A}
u^{k}_{A} f_{A},
\end{eqnarray} where all $f_{A}$'s are evaluated at ${r,q}$.  The label
{\sl density} is a misnomer, since it does not have the physical
dimensions of a density; we use it only to distinguish it from $n$ --
which we introduce below -- the size of the particle flow world vector
(which does not have the physical dimensions of a density either).
Densities can naturally appear {\sl only} in a continuum model or in
approximations to it.

Each flow can also be written as the product of a scalar and a time-like
unit world vector.  The scalar specifies the strength of the flow and the
world vector its spacetime direction:
\begin{eqnarray} N^{i} &=& n U^{i},\\ S^{i} &=& ns V^{i} .  \end{eqnarray}
This introduces the strength of the particle flow $n$, and its direction
$U^{i}$; the strength of the entropy flow $ns$; and its direction $V^{i}$.
It is important to stress that we now have two different and equally
sensible invariant definitions of a scalar quantity which qualify as a
``particle density'', the strength $n$ of the particle flow, and $\rho$,
the latter being also equal to the (negative) trace of the energy-momentum
tensor.  It is unclear which one is better suited as a variable.

Other directions and strengths are specified through the eigenvectors and
eigenvalues of the symmetric energy momentum tensor $T^{ik}$.  These are
defined through the equations
\begin {equation}
 T_{ik}A^{k} = a\eta_{ik}A^{k}. \end{equation} There are three
eigenvectors $A^{k}$, each with an associated eigenvalue
$a$.  Of these, the most important is the direction $W^{i}$ and the
strength $ne$, this being the time-like eigenvector and its eigenvalue,
giving rise to the energy flow world vector
$neW^{i}$.  The two other eigenvectors are space-like with eigenvalues
giving the pressures $P', P''$.  The two pressures are associated with
forces on surface elements with normals along the spatial projection of
the two space-like eigendirections.  The quantities $s$ and $e$ are the
invariant entropy {\sl per particle}, and the invariant energy {\sl per
particle}, using $n$ as the conversion factor between per lattice point
and per particle, as is usual in the continuum theory.  In lattice models
one may use with equal justice $\rho$ as the conversion factor, and
specify another invariant entropy (per particle) and invariant energy
(per particle) as $s^{'} = ns/\rho$, and
$e^{'} = ne/\rho$.

\section {EQUILIBRIUM CONSIDERATIONS}

According to this description, $three$ time-like unit world vectors,
$U^{i},V^{i},W^{i}$ vie in principle to specify preferred directions.  We
show now that in full thermal equilibrium there can only be {\sl one}
preferred direction, provided there is only {\sl one} set of dynamical
conservation laws to be satisfied. In general, this is the conservation
of world momenta.  (In the present model, however, one has additional
conservation laws.)  Outside thermal equilibrium this need no longer be
the case, and more preferred directions may exist.

In the absence of external fields of force we expect that the stationary
states are homogeneous in spacetime, {\sl i.e.},
$f_A$ is independent of $r$.  If this be so, and if $c_A$ vanishes,
evolution ceases; a stationary state is present.  If all possible
stationary states are simultaneously present, presumably thermal
equilibrium is reached.  (The final test is whether the entropy has
reached its maximum value given certain constraints.  This, however, we
will not investigate here but take it for granted.)

 The vanishing of
$c_A$ requires the validity of the scalar relation
\begin{equation}
 f_{A}f_{A+2}(1 - f_{A+1})(1 - f_{A+3}) = f_{A+1}f_{A+3}(1 - f_{A})(1 -
f_{A+2}),
\end{equation} or
\begin{eqnarray} &&log[f_{A}/(1 - f_{A})] + log[f_{A+2}/(1 - f_{A+2})]
\nonumber\\
&=&log[ f_{A+1}/(1 - f_{A+1})] + log[f_{A+3}/(1 - f_{A+3})].
\end{eqnarray} From this it follows that
\begin{equation} Q_{A} + Q_{A+2} = Q_{A+1} + Q_{A+3},
\end{equation} where we have renamed the arguments of the log functions
by $Q$.  One particular solution is given by
\begin{equation}
 Q_{A} = \alpha,
\end{equation} where $\alpha$ is a constant, independent of $u_A$.

Consider now the conservation of world momentum in a pair collision,
\begin{equation}
 u_{A}^{i} + u_{A+2}^{i} = u_{A+1}^{i} + u_{A+3}^{i}.  \end{equation}
Scalar multiplication of this relation with a world vector, say
$\beta_{i} $, gives
\begin{equation}
\beta\cdot u_{A} + \beta\cdot u_{A+2} = \beta\cdot u_{A+1} + \beta\cdot
u_{A+3}.
\end{equation}
 Putting $Q_{A} = \beta\cdot u_{A}$ we recover the basic relation between
the $Q_{A}$'s above.

If there are no more conservation laws, these are {\sl all} the particular
solutions.  Adding the two solutions we find that
\begin{equation} log[f_A/(1-f_A)] = \alpha + (\beta \cdot u_{A}),
\end{equation} or
\begin{equation} f_{A} = 1/[1 + exp(\alpha + (\beta\cdot u_{A}))],
\end{equation} where $\alpha$ is a scalar constant and $\beta^{i}$ is a
constant world vector.  The latter can, of course, be written as the
product of its magnitude and a unit world vector.

Further statistical considerations are needed to determine the physical
meaning and values of these parameters from the given conditions, as the
invariant density
$\rho$, (or $n$), and the invariant energy $ne$.  One expects that
$\alpha$ is proportional to the chemical potential, while the world
vector $\beta^{i}$ is related to a time-like unit world vector describing
the average flow direction in spacetime, its magnitude being the
reciprocal temperature.  The details, however, will depend on whether we
define the invariant density via $\rho$, or via $n$.  Moreover, at this
stage it is not clear whether the flux determined in this manner can be
reduced to rest through an allowed discrete Lorentz transformation, or
any combination of them.  (One may observe that the $f_{A}$ given above
makes $c_{A}$ vanish even if $\alpha$ and $\beta^{i}$ are functions of $r$
and ${q}$.  However, in this case the left hand side of Boltzmann's
equation need no longer vanish.  These ``local equilibrium functions''
are useful in studying different possible linearizations.)

 Perhaps it is useful to stress that this derivation shows why in strict
thermal equilibrium, only {\sl one} preferred direction, that of
$\beta^{i}$, can exist; thus the spacetime directions of all fluxes must
coincide.  This is the consequence of the fact that in general only {\sl
one} vectorial conservation law can exist (the conservation of world
momentum in a pair collision).  Thus its conversion to a scalar
expression requires the introduction of only one world vector.  If,
however, there are additional conservation laws, this need not be the
case.  Outside equilibrium, with irreversible processes present, new
preferred directions may arise.

An identical looking expression has been obtained in the non-relativistic
case, see \cite{Frisch}.  There, however, the scalar product appearing is
between vectors in $space$, while here between $world$ vectors.
Consequently, the non-relativistic result is not Galilean invariant
(though it should be), while the present one is Lorentz invariant.

In this simple model {\sl additional conserved quantities} exist and they
will modify the equilibrium distribution.  This occurs as follows.  In
the special frame we find

$\begin{array}{cc}
 u_{A} + u_{A +2} = \left(\begin{array}{c}6\\0\\0\end{array}\right),&
 u_{A +1} + u_{A + 3} = \left(\begin{array}{c}6\\0\\0\end{array}\right).
\end{array}
$

Thus, the line of collision of the incoming particles, and the lines of
departure of the rebounding ones lie in {\sl one} plane, the collision
plane, in which the components of the sums of incoming and outgoing
velocities vanish {\sl separately}.  Multiply the first sum with a world
vector $K$ whose time-like component is zero but arbitrary otherwise; the
second with the world vector $K'$ whose time-like component is zero but
arbitrary otherwise.  (Thus the world vectors $K,K'$ are space-like.) We
get a new particular solution
\begin{eqnarray} Q_{A}&= &K \cdot u_{A}\\ Q_{A+2}&=& K \cdot u_{A+2}\\
Q_{A+1}&=&K'\cdot u_{A+1}\\ Q_{A+3}& =& K' \cdot u_{A+3},\\
\end{eqnarray} since the sum of the first pair of $Q$'s is equal to the
sum of the second pair - as required - both being zero.  Adding this new
particular solution to the previous ones we now obtain
\begin{eqnarray} f_{A}& = &1/[1 + exp(\alpha + ((\beta + K)\cdot
u_{A}))],\\ f_{A + 2}& =&1/[1 + exp(\alpha + ((\beta + K)\cdot
u_{A}))],\\ f_{A + 1}& = &1/[1 + exp(\alpha + ((\beta + K')\cdot
u_{A}))],\\ f_{A + 3} &=&1/[1 + exp(\alpha + ((\beta + K')\cdot u_{A}))].
\end{eqnarray}

As before, further considerations are needed to give physical meaning to
the $K,K'$ vectors.  At this stage the only condition on them is that
they should be space-like, lying in the collision plane.  In the present
model, therefore, we expect that the thermalization process generates two
groups of particles, which thermalize within each group, but not the
groups with each other.  However, even the word `thermalization' is
inappropriate. Thermalization implies the existence of the notion of a
temperature which manifests itself as a spread in the energy distribution
among the particles.  Here, in the special frame the zeroth component of
all four world velocities are equal initially and stay equal during
collisions, and thus there is no spread in this frame.  Thermalization in
the present model can only mean homogenization and isotropization.

Further shortcomings of our simple model appear as well.  The {\sl
magnitudes} of the spatial velocities in this frame are all equal to each
other, being $2\sqrt2$.  Consequently, once a set of points separate from
another set of points by uniform streaming, they cannot mix again (unless
walls, or periodic conditions are applied), since no particle in one set
is capable of catching up with particles in the other set.  This latter
simplification could be rectified.

\section{NUMERICAL STUDIES}

The simplicity of the model, the associated microdynamics, and the kinetic
theory can be exploited in numerical studies.  Thus, one can study a
variety of problems both from the microdynamical and the kinetic point of
view and contrast the results. Moreover, in the latter we may use
different types of averaging methods to study the similarities and
differences.  Here we shall give a few examples of each.  (More detailed
studies are in progress, and will be reported elsewhere). (In the figures
the $x$ position
 is always counted from {\sl right to left}. This allows a simple
transcription of a $32$
$x$ positions occupancy into a binary long word that the computer uses
while iterating
particle propagation and collision.)
\vspace {0.2in}

\subsection{Frames}

 All figures, save the first, depict the evolution of systems given in a
{\sl preferred} frame in which the iteration number $q$ coincides with
the time component of the radius vector of a point.  By choosing another
frame connected with the first by a Lorentz transformation, we would
obtain a set of pictures where the points present in a {\sl t$^\prime$=
constant} plane would correspond to different iteration numbers.  The
fundamental world velocities in a general frame would appear quite
different.  Figure $1a$ shows the four fundamental world velocities from
the point of view of an observer boosted so that $u_{3}$ is given by
$(1,0,0)$, representing a particle at rest.  In this figure we see clearly
the relativistic effects.  The other three boosted
fundamental world velocities appear to move together and $u_{1}^{'}$ lies
nearly in the spacetime plane spanned by the vectors $u_{2}^{'}$ and
$u_{4}^{'}$.  (Of course it cannot lie in it, since momentum conservation
requires that $u_{1}^{'} = u_{2}^{'} + u_{4}^{'} - u_{3}^{'}$.  Since
$u_{1}^{'}$ in this frame has components $(1,0,0)$, the last term in the
sum appears to produce the smallest possible deviation from lying in this
plane.) Figure $1b$ shows how the appearance of the evolution of the four
fluids, based on the four fundamental world velocities, appear to shift
with the boost.

The subsequent figures use the preferred frame.  The points representing
particles start initially from the $t=0$ plane, and develop according to
the equations given before.  The result of each iteration therefore can
be depicted in the corresponding
$t=q$ plane.  The appropriate iteration number, $q$, is given next to each
picture.  The arrows present in the pictures indicate the spatial
directions of the four fundamental velocities.

\subsection{Examples of microscopic evolution}

On this level the relevant object to be investigated is $n_{A}(r,q)$.
This describes the actual microdynamical evolution of particles on the
spacetime lattice starting from
$n_{A}(r,0)$.  Figure $2a$ depicts the head-on collision of two sets of
points initially separated.  Each group occupies half the lattice points
of a $96 {\times} 96$ square, chosen at random {\sl elaborate}.  Within
each square the world velocities are uniform, advancing the squares
towards each other, but with different edge orientations as shown at the
zero'th iteration.  After $40$ iterations the collision of the squares is
in full bloom, and after $80$ iterations one finds four separate groups,
each containing only one particular world velocity, and streaming apart
accordingly.  The pictures show the profound differences in the
collisions generated by the interplay of the orientation -- which
regulates the arrangement of the overlap regions in the collision -- and
the preferred diagonal directions present in the collision mechanism
where the incoming and outgoing directions are along the diagonals.  The
turning mechanism in the collision can, however, only become effective if
the outgoing diagonal connects sites not yet occupied, thus depending on
the overlap.

In edge-on collisions there is a large overlap region at the start of the
collision which generates a spillout along the diagonal direction.  This
liberates sites to allow the turning process in the collision to operate
well, generating increasing spillout, and so on.  Finally we see the four
well separated final regions.  (We have briefly discussed at the end of
Section $7$ why this division into four, well separated, free streaming
regions is the consequence of the model.)

In corner-on collisions the transverse flow is confined because in this
configuration the initial target area is diminished and, in addition, the
new incoming points impede the turning mechanism.  Eventually the
collisions do produce points with new world velocities, and the system
again breaks up into four final parts, each homogeneous in the world
velocities.  (It is amusing to see how some of the sharp edges are
preserved.  This arises because during the collision of the squares these
edges advance as a common front in each step, covering up precisely the
new points spilling out due to the collision.)

Figure $2b$ shows the same arrangement in a four square collision.  Here
the large degree of symmetry does not enable one to distinguish the final
spillout regions from the others.  Moreover, in the corner-on collision
the conspiracy of the symmetries in the diagonals and edges finally
generate only an exchange of world velocities in the diagonally opposite
regions.

One may investigate how the occupancy in the initial configurations
influences these results.  Figure $3a$ demonstrates the effects of the
initial occupation density on the post-collision configurations ($q=80$)
in the two square collisions.  The left column refers to edge-on
collisions, the right column to the corner-on collisions. In the first
row the occupation fraction in the initial squares was $1/3$, in the
second row $2/3$, in the third row $1$.

Figure 3b is arranged in an identical fashion, showing the effects of
occupancy in a post collision configuration ($q=80$) for four square
collisions.  An interesting feature is the influence of the occupancy in
the edge-on collisions.  For very small occupancy the trailing regions
are small (since for single occupancy this region must vanish).  For full
occupancy all edges are sharp, no trailing region exists.  Consequently,
there must exist one or more partial occupancy numbers for which the
trailing region is a maximum.

\subsection{Kinetic examples}

\vspace{0.2in}

There are many different questions and problems one can investigate using
the kinetic approach.  Some of these we shall mention later.  Here we
take up only one special case.

Figure $4a$ shows the microscopic evolution of points in a rectangle,
i.e., the functions
$n_{A}(r,q)$.  Initially $1/8$ of the locations are occupied at random by
each of the four different world velocities, also chosen at random.  Thus
half the points in the initial rectangle are occupied.  After ten
iterations the square increases in size, and nine regions can be
distinguished in it, according to the number of different fundamental
world velocities in it.  The central region has points with all four
velocities present; such a region we call phase $III$.  The regions at
each four corners is phase $I$, containing points with only one of the
four basic velocities.  These pure regions are connected by regions of
phase $II$ with differing sizes, where the adjoining velocities are mixed
pairwise. These
regions separate during the evolution and the mixed regions progressively
disappear.  (Phases $III$ and the small phase $II$ disappear at
$q=16$, while the large phase $II$ disapears at $q=32$.)

 We now conceive an {\sl ensemble} where each member has different initial
data chosen randomly in the same manner.  Different averages over the
ensemble will generate different distribution functions.  From any one of
them we can construct all the relevant average quantities discussed
before.  Hence, in principle, we can compare the different quantities
obtained.  There are the following quantities to be compared:

a) $n_{A}(r,q)$, the solution of the microscopic equations, advancing the
initial data $n_{A}(r,0)$;

b) $f_{A}(r,q)$, the solution of the kinetic equation, advancing the
initial data obtained by averaging $n_{A}(r,0)$;

c) $F_{A}(r,q)$, the average of $n_{A}(r,q)$.

The quantities $n_{A}(r,q)$ give the complete microscopic description.
(In the present case this is given in Figure $4a$ for one particular
member of the ensemble).  The functions $F_{A}(r,q)$ give the most
correct {\sl average} description, through averaging the precise
microscopic description for each member of the ensemble at the iteration
$q$.  The functions $f_{A}(r,q)$ give the {\sl kinetic} description.
Here, following Gibbs, we average over the {\sl initial data}
$n_{A}(r,0)$ within the ensemble, and then evolve this average via
Boltzmann's equation.  (One could also use an initial average of one
particular member of the ensemble over given ``small, but not too small''
spacetime regions, following Boltzmann's ideas, and pursue the evolution
of this distribution function.  This we shall not do here.)

 In Figures $4b$, and $4c$ we show the behavior of some of these
statistical averages computed in this manner.  In principle, all ensemble
averages still depend on the location $r$, and iteration number $q$.  To
simplify our pictures we will eventually also perform an additional
averaging over sites occupied at each iteration.  To gain further insight
we shall do this averaging in two stages.  First we average over the
occupied sites in each phase separately.  These averages will be denoted
by the square brackets, $[{\cdot}]$.  Then we average over all the
occupied sites by using the averages over the fluid in this phase and
multiplying each with the fraction the phase was represented in the total
number of occupied sites.  These averages will be denoted by the pointed
brackets, $<{\cdot}>$.

In figure 4b there are several average quantities shown as a function of
the iteration number $q$.  The solid curve indicates the quantity
averaged over phase $III$, the dashed curve an average over phase $II$,
the dotted curve an average over phase $ I $. (The solid and dashed lines
terminate at those iterations where the overlap generating the phase
ceases.) The last graph in the figure shows the fractional occupancy. The
dashed line shows the fractional occupancy in Phase $III$.  There are two
types of Phase $II$ regions, large ones and small ones.(See, {\it e.g.},
Figure $4a$, $q=10$.) The small overlap regions disappear after 16
iterations, the large one after 32 iterations. Their fractional occupancy
is shown by the dashed and by the dashed-dotted curves respectively.
(Since both the large and small Phase $II$ regions contribute equally to
the overlap region in the $[{\cdot}]$ averages no distinction had to be
made between these averages, hence no dashed-dotted lines appear.)

 We notice that all average quantities jump immediately, with one
iteration, to a stationary value.  This arises since the initial values
were so random that the phases immediately reached thermal
equilibrium; only their mixing ratios could change during evolution.

Figure $4c$ shows as a function of $q$, the {\sl average magnitudes} per
occupied site of the two differently defined invariant particle numbers,
$<\rho>$, $<n>$, the
entropy and energy flows  $<ns>$,$<ne>$, and the site-averaged pressures
in the
$x$ and $y$ directions, $<P_x>$, $<P_y>$.  (In this case the associated
space-like eigendirections are along the $x$ and $y$ directions.) All
quantities are averaged over the contributing sites, including all the
different phases together.  For these quantities, in this particular
case, there are no differences visible whether we use the $f$ functions or
the $F$ functions, showing the good quality of the kinetic approximation.

The situation is somewhat different for the {\sl normalized directions} of
the flows. Instead of plotting them we shall only plot in Figure $4d$ the
magnitudes of the squared differences.  (These are symmetric in their
arguments.) For example, the difference between the particle flow
direction
$U^{i}$ and entropy flow direction
$V^{i}$ is defined as $ - (U^{i} - V^{i})(U_{i} - V_{i}) = 2(1
+ U^{i}V_{i}) $, with
$(U^{i}U_{i} = V^{i}V_{i}= - 1) $.  This difference (averaged) is denoted
by $<\delta (s,n)>$, or $<\Delta (s,n)>$, depending whether we evaluate
it using the $f$, or $F$ distribution functions.  The other differences
are similarly defined.  The labels $n,s,e$ in the figures refer to the
associated flows.

\subsection{Equation of state}
 Figures 4b and 4c have shown values for
the pressures on planes normal to the two spatial directions.  Figures
$5a,b$ show typical graphs used in the construction of an equation of
state associated with this model.  Figure 5a uses the distribution
function $f(r,q)$, while 5b uses the function $F(r,q)$.    The two pressures
$P_{x}, P_{y}$ and the energy density $ne$ are thus functions of $r$ and
$q$.  Here we see their relation at different lattice points at the $q=5$
iteration.  The values of both pressures are indicated on the vertical
axis, the energy density on the horizontal one. In each picture four
clouds of points appear associated with the regions of different phases
(velocity mixtures).   The small cloud at the origin depicts pure velocity
phases, where both pressures vanish. (On the scale of the figure this cloud
is invisible.)  The
large cloud has all phases present.  The two regions where only {\sl two}
velocities mix are
disjoint, since in these regions the energy densities are the same, but
$P_{x}$ is zero
while $P_{y}$ is not.  These regions have a finite size due to the finite
number of initial data
in the ensemble.  These sizes also depend whether we use $f(r,q)$, or
$F(r,q)$ in the averaging,
since each distribution function gives rise to different fluctuations.  As
the number of initial
data increases, the different regions reduce in size, converging towards
four points, given a
density.  Consequently, varying the density we finally generate four lines
giving the equation of
state. (Two of these line coincide, giving $P_{x} = 0$ for both.)
We thus obtain

\begin{eqnarray} P_{x} & = & 0,\\
 P_{y} & = &(4/5)ne,\\
\rho & = & (1/5) ne,
\end{eqnarray}for the two velocity region;
\begin{eqnarray} P_{x} & = & P_{y} = \nonumber\\
P & = & (4/9)ne,\\
\rho & = & (1/9)ne,
\end{eqnarray} for the four velocity region;
\begin{eqnarray} P_{x} & = & 0,\\
P_{y} & = & 0,\\
\rho & = & ne,
\end{eqnarray} for the free streaming regions.

\section{OUTLOOK AND FURTHER QUESTIONS}

The above discussion and examples show that a relativistic lattice
automaton describing relativistic fluid flows can be constructed.  One is
able to confront {\sl directly} the microdynamics, {\it i.e.},  the actual
relativistic evolution with the one computed from the kinetic model.
This may enable us to investigate both general and particular questions.
Some of these are listed below.

a) To proceed, more {\sl realistic} models and initial data are needed.
We may improve the collision mechanism in order to destroy the additional
constants of the motion, and to generate a spread in the particle
energies.  This will require the use of more than four fundamental world
velocities, and, possibly, different spacetime lattices as well.

b) Given this, we can evaluate in more appealing models the various
quantities containing sums over $A$, using the three different
distribution functions ($n_{A},f_{A},F_{A}$) and compare them with each
other to study the quality of the approximations involved in using
averages, and using the kinetic equation (as we have done briefly in
figure 3c).  This may show the validity and range of the different
assumptions entering the kinetic theory and its linearized approximations;
it can also exhibit the fruitfulness (or otherwise) of the different
definitions used. For example, what time-like unit vector should be
chosen (if any) to describe the ``hydrodynamical flow''; whether $n$, or
$\rho$ should be used as the invariant occupation number on a site, etc.

c) Of special interest are questions connected with the entropy.  The {\sl
microscopic} entropy flux is identically zero, because the logarithmic
terms identically vanish (on account of $n_{A}$ being either unity or
zero).  This vanishing is indeed appropriate, the entropy being a
statistical notion.  There remain, however, still two statistical entropy
flux definitions, using either the $F_{A}(r,q)$, or the $f_{A}$ functions!
The one defined through $F$ is the actual entropy flux, the other is its
kinetic approximation.  How will these two different entropy fluxes
differ, and how will they result in a different entropy production?

 The entropy production itself is an important problem.  We mention here
two particular classes of questions.

For the entropy defined through the $f$ functions, the Boltzmann equation
insures the existence of an H theorem (for confined systems, or infinite
systems) on the {\sl kinetic} level.  There exists, however, in this case
no microscopic entropy flux using the $n$ functions, and therefore no H
theorem.  What will then happen if we compare the long term solutions
(large iteration numbers) of the microscopic equation, via an entropy
defined through the $F$ functions, ({\it i.e.}, the correctly evolved,
averaged $n$ functions), with the equilibrium solutions? Disparities
should exist due to the approximate nature of the kinetic evolution.
Boltzmann himself conjectured long ago that fluctuations of the true
entropy should arise even in equilibrium.  Will the entropy defined
through $F$ verify this conjecture? We expect so.

All irreversible processes are linked to entropy production.  To specify,
however, {\sl macroscopic} irreversible processes linked to this entropy
production, macroscopic irreversible fluxes and its driving ``forces''
must be introduced.  These are eventually connected by coefficients,
displaying symmetry properties, the Onsager relations, exhibited by {\sl
macroscopic} time reversal arguments \cite {Wigner}.  In conventional
relativistic hydrodynamics this step and its consequences generate great
difficulties, and contradictory results.  The usual {\sl kinetic }
approach, based on the solutions of the kinetic equations near
equilibrium has not resolved the difficulties.  Another basic problem is
to find the correct {\sl relativistic} Onsager relations, which would
then lead to hyperbolic equations describing the irreversible processes.

It seems to be essential to find an {\sl invariant} small parameter.
This is where much of the previous work, imitating Chapman and Enskog in a
relativistic context, has difficulties.  We conjecture that there exists
a {\sl mean free iteration number} corresponding to the mean free time in
the non-relativistic case.  The small parameter may then be the ratio of
the iteration number over the mean free iteration number, or some other
quantity related to them.

d) Conservation Orbits

The correct solutions of the kinetic equations can generate additional
conserved quantities.  For example, $\rho(r,q) = \sum_{A} f_{A}(r,q)$ is
such a quantity, implying that $\sum_{A} f_{A}(r,q) =\sum_{A} f_{A}(r +
u_{A},q + 1)$.  In other words there can be a point $r'$, (or points)
such that $\rho(r',q + 1) = \rho (r,q)$.  If there is only one such point
at each iteration the successive points generate an orbit in spacetime.
(These points need not be unique; for example if $f_{A}$ is independent
of $r$, as in equilibrium, there will be an infinite number of them.)
With each conserved quantity a conservation orbit could exist, but need
not.  If they do, different time-like ``tangent'' vectors can be defined
by taking two adjacent points on an orbit, referring to iteration $q$ and
$q-1$, taking their difference and normalizing them to $-1$.  What is
their relation to the other unit time-like vectors specified through the
fluxes?

\section {SUMMARY}

We have shown how one can construct spacetime lattices using the discrete
subgroups of the Poincar\'e group, and define over them a dynamics that
is as relativistic as possible replacing a continuous group of
transformations with a discrete one.  Then, using the simplest example we
exhibit such a dynamics with contact interactions at the lattice sites,
using both a detailed microdynamics and its kinetic approximation based
on an associated Boltzmann equation.  Apart from some analytical results
we also showed numerical ones representing simple collisions.  Finally,
we have offered some questions for future work.

\section{APPENDIX}

One can establish a formal correspondence between the discrete formalism
and the continuum one using the formally invariant relations
\begin{eqnarray}
\sum_{A} &\longrightarrow& \int\! d^{2}p/p^{0}\\ r &\longrightarrow&
x^{i}\\ u^{i}_{A} &\longrightarrow& p^{i}\\ f_{A} &\longrightarrow&
f(p,x)\; .  \end{eqnarray}

\pagebreak

\newpage

\section{Figure Captions}

Fig. 1a The appearance of the four fundamental world velocities $u_{A}^{'}
= L(1) u_{A}$ in the frame obtained from the special frame by the boost
$L(1)$. Fig. 1b The spacetime evolution of four fluids in the special
frame (lower plot), and in the boosted frame.

Fig. 2a. History of edge-on and corner-on collisions of two squares
populated by particles, as a function of iteration number $q$. (Detailed
description in text.)

Fig. 2b. History of edge-on and corner-on collisions of four squares,
populated by particles as a function of iteration number $q$. (Detailed
description in text.)

Fig. 3a. Post-collision situations ($q = 80$) in edge-on and corner-on
collisions of two squares, populated by particles with varying fractional
occupancy, as indicated on the top of each picture. (Detailed description
in Section VIII.)

Fig. 3b. Post-collision situations ($q = 80$) in edge-on and corner-on
collisions of four squares, populated by particles with varying
fractional occupancy, as indicated on the top of each picture. (Detailed
description in VIII.)

Fig. 4a. History of collisions of particles situated originally in a
rectangle as a function of iteration number $q$. (Detailed description in
Section VIII.)

Fig. 4b. Averages of density $[\rho ]$, magnitude $[n]$ of particle flow,
magnitude $[ne]$ of energy flow , magnitude $[ns]$ of entropy flow,
principal pressures $[P_{x}]$, $[P_{y}]$. These averages in regions
$III,II,I$, denoted by $[{\cdot}]$ are restricted to sites in the different
regions
associated with different phases. (Detailed description in Section VIII.)

Fig. 4c. Average magnitudes for quantities in Fig. 4b, but averaged over
all contributing site areas, irrespectively of the particular phase,
denoted by $<{\cdot} >$. (Detailed description in Section VIII.)

Fig. 4d. Average deviation between pairs of flow directions indicated.
Deviations $\delta$ are evaluated using the kinetic distribution function
$f$; deviations $\Delta$ are evaluated using $F$, the microscopic
distribution function averaged. The averages, denoted by $<{\cdot}>$ are over all
contributing site areas with deviations different from zero. (No such
deviations can, of course, in the one fluid phase region. Detailed
description in Section VIII.)

Fig. 5a. Graph showing relation between pressures and energy densities at
different sites for iteration $q = 5$ using function $f$. (Detailed
description in Section VIII.)

Fig. 5b. Showing same, using function $F$. (Detailed description in
Section VIII.)

\end{document}